\documentclass{article}
\usepackage[preprint]{spconfa4}
\usepackage{amsmath,graphicx,enumitem}
\usepackage{makecell}
\graphicspath{ {./images/} }
\copyrightnotice{978-1-6654-6867-1/22/\$31.00~\copyright2022 IEEE}
\usepackage{array}
\title{Mechatronic generation of datasets for acoustics research}

\customname{{Austin Lu, Ethaniel Moore, Arya Nallanthighall, Kanad Sarkar, Manan Mittal,}
\thanks{This work was supported in part by The Discovery Partners Institute and the Grainger College of Engineering at the University of Illinois Urbana-Champaign. The work of R. Corey was supported by an appointment to the Intelligence Community   Postdoctoral Research Fellowship Program at the University of Illinois Urbana-Champaign, administered by Oak Ridge Institute for Science and Education through an interagency agreement between the U.S. Department of Energy and the Office of the Director of National Intelligence.}}
     {Ryan M. Corey, Paris Smaragdis, and Andrew Singer}
\address{University of Illinois Urbana-Champaign}

\begin{document}
\maketitle
\begin{abstract}
We address the challenge of making spatial audio datasets by proposing a shared mechanized recording space that can run custom acoustic experiments: a Mechatronic Acoustic Research System (MARS).
To accommodate a wide variety of experiments, we implement an extensible architecture for wireless multi-robot coordination which enables synchronized robot motion for dynamic scenes with moving speakers and microphones. Using a virtual control interface, we can remotely design automated experiments to collect large-scale audio data. This data is shown to be similar across repeated runs, demonstrating the reliability of MARS. We discuss the potential for MARS to make audio data collection accessible for researchers without dedicated acoustic research spaces.

\end{abstract}
\begin{keywords}
Audio recording, robotics, remote control, signal processing, source separation, speech processing
\end{keywords}
\section{Introduction}
\label{sec:intro}

Rich, high-quality audio data is a vital resource for the development and evaluation of audio signal processing algorithms. However, the process of acquiring data can be time-consuming and labor-intensive.

Researchers have developed a number of approaches for generating audio data with desired spatial characteristics. Simulation methods such as the Image Source Model (ISM) can be used to generate a room impulse response (RIR), which can be convolved with an audio signal to model room acoustics \cite{ISM}.
Unfortunately, RIRs are hard to leverage in live applications. Therefore, researchers interested in evaluating their algorithms in live settings will use data from carefully arranged microphones and loudspeakers \cite{ryan}. 
Some researchers opt to record human subjects engaging in everyday tasks, which provides rich, dynamic data \cite{CHiME, REVERB}. Sensors can be used to measure the position of subjects \cite{wargames}, which is useful for research in localization and tracking. Although realistic, these experiments are difficult to replicate.
In contrast, robotic experiments are repeatable and can emulate complex human-like motion. An example is the LOCATA challenge, which provides a data corpus that includes recordings from a moving humanoid robot \cite{locata1, locata2}. Robots also enable dense spatial samplings of acoustic spaces. The CAMIL dataset was created for research on binaural manifolds \cite{CAMIL} by using a robot to direct an acoustic head simulator at various orientations. Although robots can benefit spatial audio research, they are costly to develop and thus inaccessible to many.

\begin{figure}[t]
	\centering
	\centerline{\includegraphics[width=7cm]{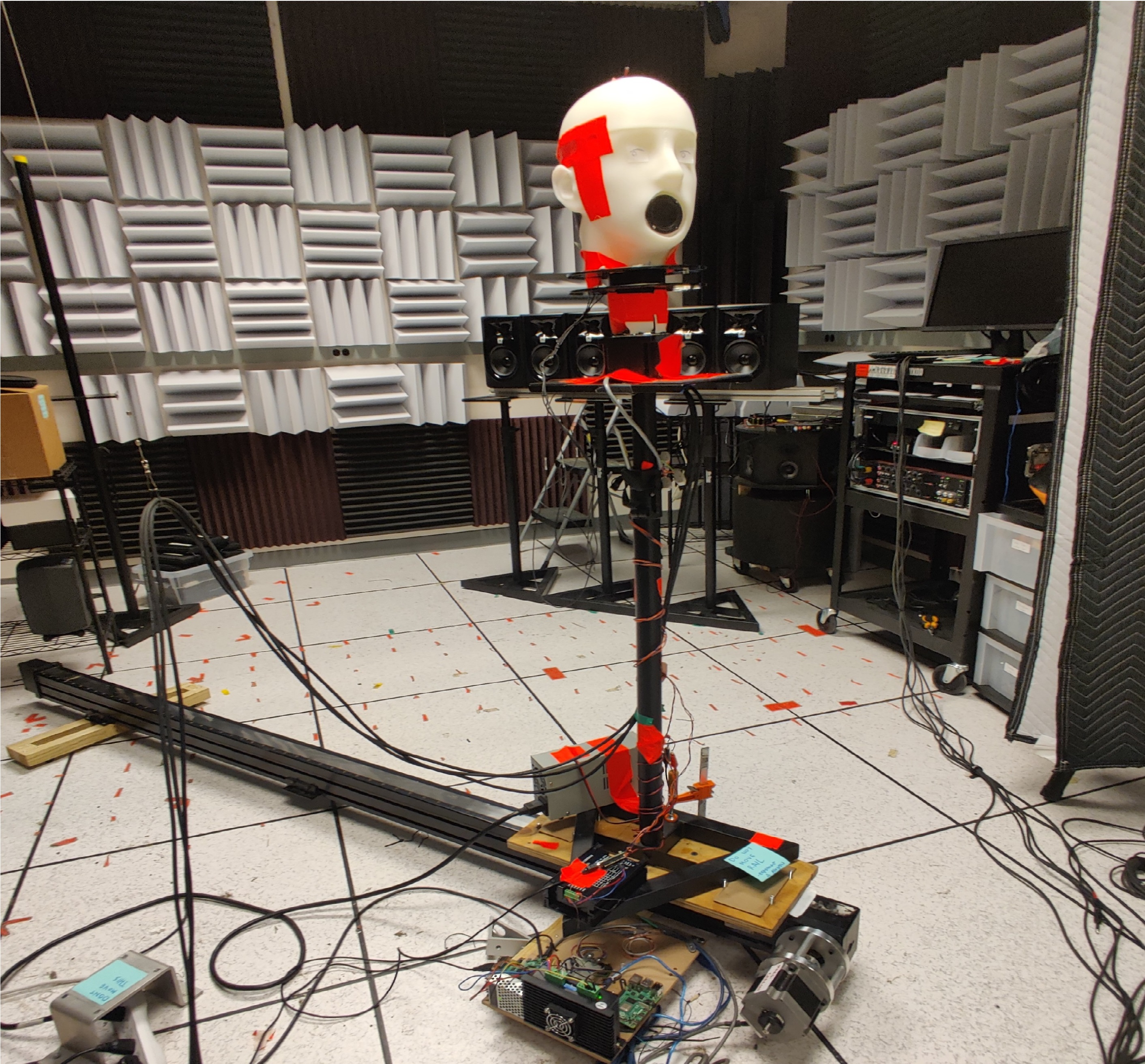}}
	\caption{Photograph of the MARS setup used in this paper}
	\label{fig:room}
\end{figure}

A tool that grants access to a robot-enabled recording space would enable many researchers to generate real data from finely controlled acoustic scenes.
The Mechatronic Acoustic Research System (MARS) is such a  remotely-accessible workbench for the creation of audio datasets.
We summarize the tradeoffs of our proposed system with other approaches in Table \ref{tbl:summary}.

\begin{table}[htbp]
    \begin{center}
        \resizebox{\columnwidth}{!}{
        \begin{tabular}{ | p{1.8cm} | p{1.8cm} |p{1.8cm} |p{1.8cm} |p{1.8cm}|} 
            \hline
            & Simulation & Live human experiment & Live robotic experiment & \makecell[lt]{MARS\\prototype} \\ [0.5ex] 
            \hline
            \hline
            Modeling Inaccuracies & Inexact \cite{pra, dschr} & Inexact Positioning and Labels &Servo Noise from Robotic Motion& Servo Noise from Robotic Motion \\
            \hline
            \makecell[lt]{Required \\ Resources} & Low & High & High & Remotely Accessible \\
            \hline
            Repeatable & Yes & No & Yes & Yes \\ [1ex] 
            \hline
            Setup Time & Minimal & High &  High & Minimal\\
            \hline
            \makecell[l]{Dense spatial \\ sampling} & Yes & No & Yes & Yes\\
            \hline
        \end{tabular}}
    \end{center}
    \caption{Summary of spatial audio experiment methods.}\label{tbl:summary}
\end{table}

The challenges that must be solved for MARS to achieve our goals are explored in Section \ref{sec:challenges}. Design and control of the MARS prototype is discussed in detail in Sections \ref{sec:design} and \ref{sec:control}. We evaluate our prototype's data-collection ability in Section \ref{sec:eval} and summarize our results in Section \ref{sec:conclusion}.

\section{Challenges}
\label{sec:challenges}

For MARS to be a viable general-purpose audio data collection tool, it must allow users to specify a wide variety of experiments and be able to run them consistently. We refer to these two tasks as design and control, respectively.
To design an experiment, a user must be able to fully describe complicated acoustic scenes, which may involve many speakers and microphones, each with their own audio, positioning, and motion. The more elements in an experiment, the more challenging it is to design. To address this, MARS must represent acoustic scenes in a concise manner, such that users can easily specify even the most elaborate of experiments.
Additionally, MARS is intended to be remotely accessible, therefore visualization tools are necessary for users to fully understand the recording space's capabilities and limitations.

The control aspect of MARS also presents many unique challenges.
Multi-robot coordination is required for repeatable motion across a fleet of robots, however this is a complex task \cite{mrs}, and MARS must address this while also being extensible so that state-of-the-art devices can be seamlessly integrated on a rolling basis.

Furthermore, certain precision constraints must be met for MARS to be suitable as a shared data-collection platform.

\section{User design of complex experiments}
\label{sec:design}

\subsection{Describing an experiment}
Central to the task of designing an experiment is the ability to accurately specify an acoustic scene. The volume of information required to fully characterize such scenes makes concise representation challenging, thereby increasing the tedium of creating and iterating experiment designs. Our solution is to provide an object-oriented application programming interface (API) in Python, modeled loosely after \emph{pyroomacoustics} \cite{pra}. 
Any audio or robot device in MARS is dubbed a \texttt{component}, each of which offer a range of \texttt{actions} that can be requested via \texttt{instructions}. A first pass of the experiment can be done without calling upon the actual equipment, so that logs are provided nearly instantly. We find that this streamlines the process of iterating experiment designs.

\begin{figure}[t]
	\centering
	\centerline{\includegraphics[width=8.25cm]{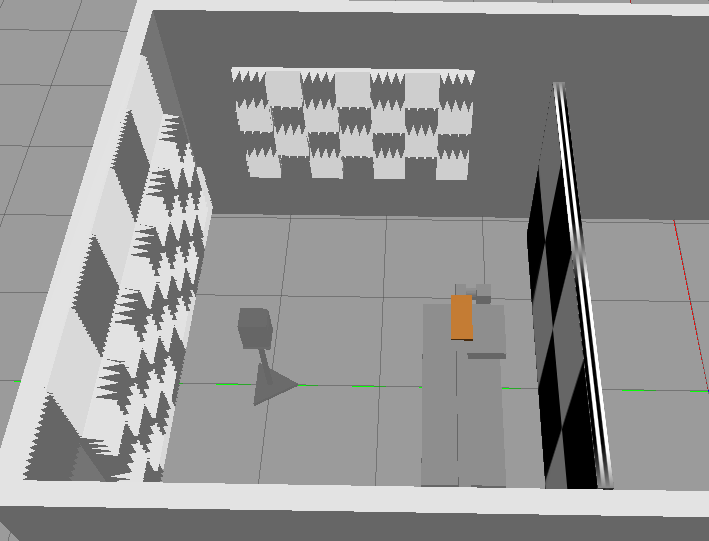}}
	\caption{A Gazebo simulation of the prototype recording space}
	\label{fig:gazebo}
\end{figure}

\subsection{Simulation in Gazebo}
Remote access to a mechanized room poses several safety concerns. In particular, the possibility for user-defined experiments to damage equipment must be addressed. A \textit{digital twin} of the system is developed in Gazebo: a simulation toolbox and physics engine commonly used in robotics \cite{gazebo}, such that experiments may be simulated in advance to check for collisions. This tool also provides a visualization of the physical space, as shown in Figure \ref{fig:gazebo}. We anticipate this tool to be instrumental in the design and execution of a new class of robot-enabled audio experiments.

\subsection{Designing and monitoring}
\label{ssec:monitor}
We operate the MARS prototype using a virtual interface, which lets us design and run experiments remotely. Experiments may involve a large number of devices, or have a duration spanning multiple days, thus a monitoring tool is provided to track experiment progress, report system status, and stream a camera feed of the recording space.

\section{Control of audio and robot devices}
\label{sec:control}

\subsection{Multi-robot coordination}
To convert descriptions of experiments into audio data, MARS must offer positioning, playback, and recording capabilities. Our prototype does this by providing an extensible API that defines such behavior for an arbitrary set of devices.

The positioning of microphones and loudspeakers within an acoustic scene is done using multiple robots, which can range from simple linear actuators to complex pulley-driven systems. These robots must move to requested positions at specified times while avoiding collision. Collision avoidance in multi-robot systems is well studied \cite{mmrs} and can be highly sensitive. Fortunately, MARS controls a known environment where motion planning can be done in advance. Additionally, for the sake of repeatability, the motion provided by MARS is mostly described by the user, so there is little need for advanced decision-making and path-planning algorithms. These factors make MARS suitable for a centralized architecture, which can be more efficient than a decentralized one \cite{efficientmrs}.

\subsection{Device integration}
\begin{figure}[t]
	\centering
	\centerline{\includegraphics[width=8.5cm]{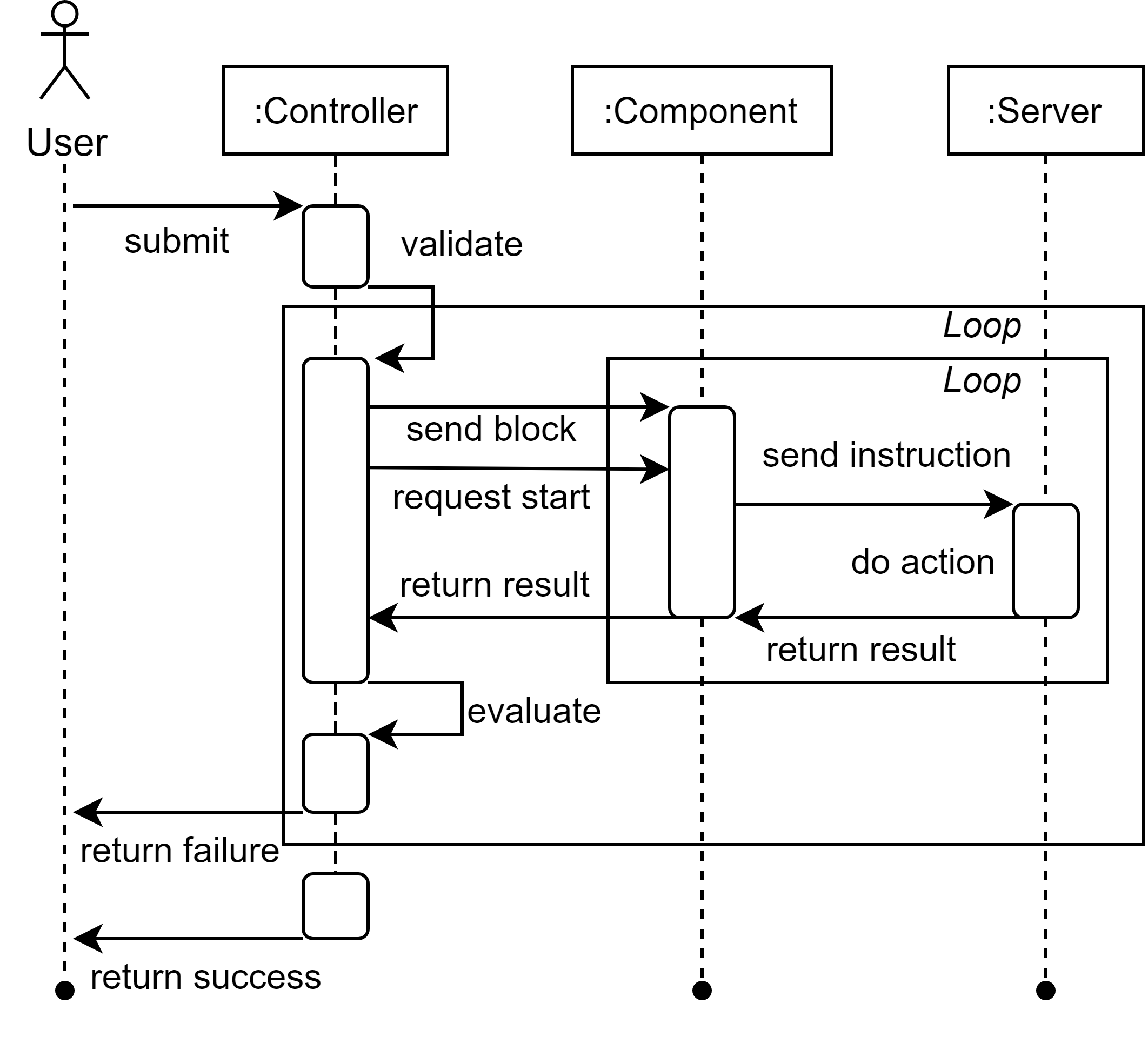}}
	\caption{Sequence diagram of MARS experiment}
	\label{fig:sequence}
\end{figure}

We acknowledge the need for MARS to provide access to state-of-the-art equipment. To this end, MARS is built atop the Robot Operating System (ROS): a collection of open-source robotics software that offers a publisher-subscriber communication architecture over IP \cite{ros}. To support device integration with MARS, we offer modular client and server objects that request and execute \texttt{instructions}, respectively. The use of client/servers in the handling of an experiment is shown in Figure \ref{fig:sequence}. A single \texttt{component} can run many servers in parallel, each of which passes instruction data to an arbitrary callback function. This implementation is extensible: Updating a server to fit a new device only requires changing the instruction type and callback implementation, so devices relevant to audio research such as the Tympan \cite{tympan} can be added with little overhead. Using ROS, wireless devices can be interfaced easily, making MARS suitable for research on cooperative listening with IoT devices. 

\subsection{Precisely timed actions}
Although the modified TCP protocol provided by ROS guarantees data integrity and order of arrival, large latency and jitter is observed, particularly on the low-power microcontrollers that drive many of the robots within MARS. This does not pose an issue for sequentially executed instructions, as order of execution is easily maintained. To support timestamped instructions, the prototype controller transmits messages in advance, which gives components ample time to receive and execute instructions according to their internal hardware clocks. Assuming synchronized hardware clocks across the network, this guarantees that robot motion occurs on a shared timeline. The effects of clock drift are minimized by running a local Network Time Protocol (NTP) server \cite{ntp}. This approach allows for the creation of dynamic acoustic scenes, even on busy wireless networks.

\subsection{Precise motion}
Given that MARS is designed to provide spatial audio, we aim to record and validate the positions of robots in the environment. Recording this data can be achieved using ROS tools, namely rosbag.
Validation requires ground-truth position data, which can be found using a camera system and visual fiducial markers \cite{visualmain, visual}.
With the ground-truth positions known, position error can be calculated and used to interrupt experiments that cause the MARS system to behave erratically. Doing so serves as a form of quality control, wherein only high-quality, precise data is captured.

\subsection{Managing large datasets}
Complex experiments have the potential to involve a massive number of devices, each of which logs its own performance locally and writes audio files to internal storage. MARS offers the ability to consolidate this data onto the host machine, where it can be labeled and uploaded for the user to access. Data upload can be requested concurrently to experiment execution, allowing for validation and feature extraction even as an experiment is ongoing. This minimizes the downtime between requesting an experiment and receiving data.

\section{Evaluation}
\label{sec:eval}
Data collected from baseline experiments was used to evaluate MARS. The following equipment setup was used:

\begin{itemize}[noitemsep,topsep=0pt]
  \item The \texttt{interface} component, a fanless computer with a 64-channel audio interface (Antelope Audio Galaxy 64 Synergy Core) allows an audio file to be played through a speaker array while a set of omnidirectional lavalier condenser microphones records audio concurrently. Scripting is handled with PortAudio \cite{portaudio} linked to the interface's ASIO driver.
  \item A 3D-printed acoustic head simulator, shown in Figure \ref{fig:room}. One microphone was placed atop the head, and two were inserted into the left and right ear canals.
  \item The \texttt{spiderbot} component, modeled after the spider-cam system used in professional sports arenas, which carries a speaker along a three-dimensional path.
  \item The \texttt{turret} component, which rotates the head.
  \item The \texttt{rail} component, a motor-driven linear guide rail that translates the \texttt{turret} along a single axis.
\end{itemize}

\subsection{High density spatial sampling}
\label{ssec:dense}
MARS was used to take a dense spatial sampling of an acoustic scene, to demonstrate robustness. The acoustic head simulator was placed in various poses by the \texttt{rail} and \texttt{turret}. Over the course of around fifty hours, 40,000 three-channel audio files were collected at a sampling frequency of 48 kHz without human supervision or intervention.

\subsection{Repeatability of experiments}
\label{ssec:repeatable}
\begin{figure}[t]
	\centering
	\centerline{\includegraphics[width=8.5cm]{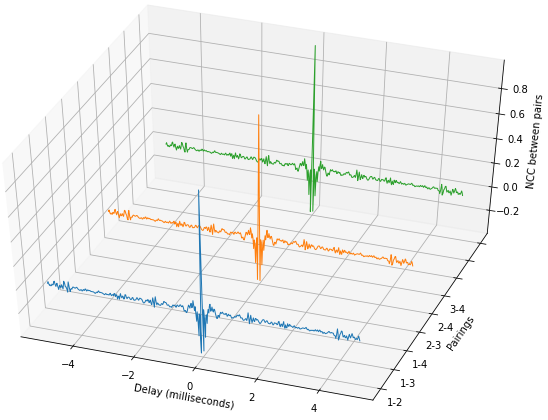}}
	\caption{Three cross-correlations corresponding to the same pose. The amplitude of these peaks are used as indicators of similarity between pairs of recordings}
	\label{fig:eval}
\end{figure}

Using the densely sampled MARS data, we verify the repeatability of MARS for static scene creation using four measurements corresponding to the same head pose. The recorded data was highpass filtered to remove background noise.
The value of the maximum entry in the normalized cross-correlation (NCC) between two signals is used as a metric for similarity between recordings. To evaluate repeatability for multiple instances of the experiment, the NCC between the first recording and each of the others is calculated, as shown in Figure \ref{fig:eval}. The maximum entries of the NCCs had an average value of 0.98 with a standard deviation of 0.01. A maximum NCC value of 1.00 corresponds to perfect similarity between two signals, thus our results indicate highly repeatable performance.

To verify repeatability for dynamic scenes, an experiment was run where the \texttt{spiderbot} carried its speaker payload along a trajectory as audio was being played concurrently. Recordings were taken from a fixed microphone across ten repeated runs of the experiments. Across nine comparisons to a reference recording, the maximum NCC was 0.93 on aver age, with a standard deviation of 0.04. We observe that the repeatability was only marginally inferior when motion was introduced.

The dynamic experiment was repeated with the fixed acoustic head simulator in place of the microphone. The interaural time difference (ITD) remained constant across separate runs of the experiment, demonstrating reliable collection of spatial audio.

\section{Conclusion}
\label{sec:conclusion}
MARS demonstrates how remote access to coordinated robots can be applied to the collection of custom high quality audio data. Using a virtual interface, we ran several experiments which would have otherwise required significant human labor and time. We evaluated this data to show that our framework for wireless multi-robot coordination is capable of collecting repeatable data from both large-scale static scenes and challenging dynamic scenes. The process of designing experiments was streamlined by our scripting interface which concisely describes the motion and playback/recording of robot-driven microphones and loudspeakers. By creating MARS with extensibility in mind, we open up the possibility of accommodating an ever-growing variety of experiments. With the development of this initial version of MARS, we have provided solutions to several of the major challenges that must be solved for an open-access mecha-acoustic platform to become a reality. 

\begin{figure}[t]
	\centering
	\centerline{\includegraphics[width=8.5cm]{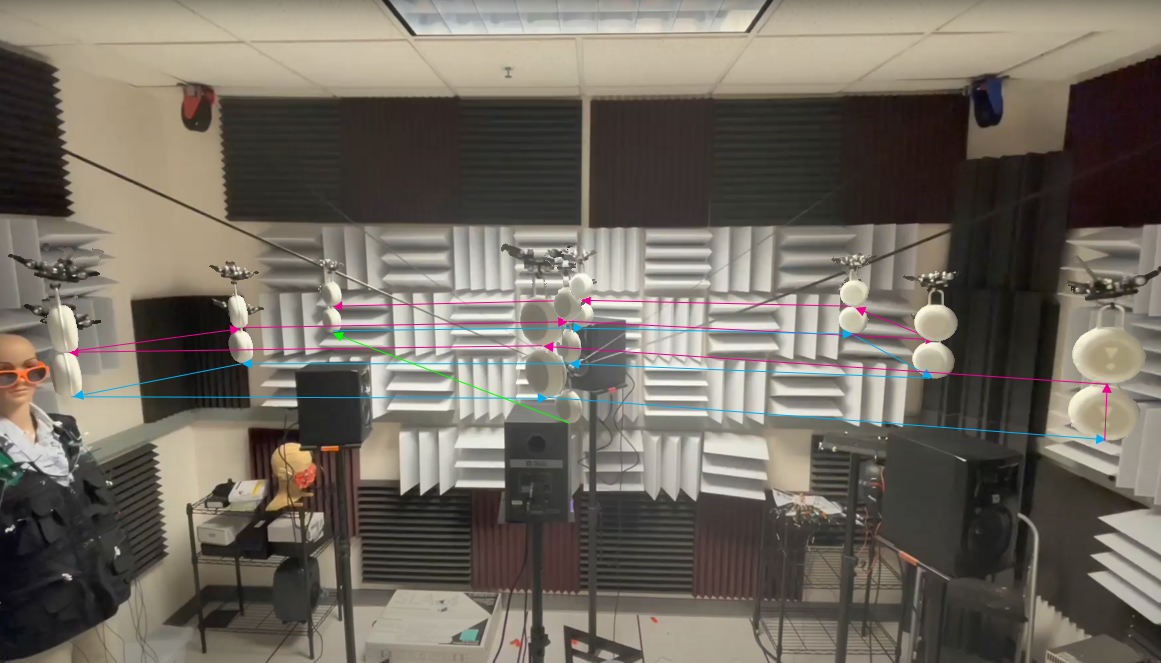}}
	\caption{Superimposition of the \texttt{spiderbot} at various timesteps along a trajectory}
	\label{fig:trajectory}
\end{figure}

\bibliographystyle{IEEEbib}
\bibliography{refs}

\end{document}